\documentclass[aps,prl,groupedaddress,superscriptaddress,twocolumn,10pt,longbibliography]{revtex4-2}
\pdfoutput=1 
\usepackage[utf8]{inputenc}
\usepackage{hyperref}
\usepackage{graphicx}
\usepackage{setspace}
\usepackage{orcidlink}
\newcommand{\corauthor}[2]{
    \author{#1}
    \email{#2}
}

\usepackage{siunitx} 
\usepackage[caption=false]{subfig}
\usepackage{tabularray}
\usepackage{amssymb}
\usepackage{physics}
\usepackage{float}
\DeclareSIUnit\bohr{\text{\ensuremath{a_\textup{0}}}}
\newcommand{\cev}[1]{\reflectbox{\ensuremath{\vec{\reflectbox{\ensuremath{#1}}}}}}
\begin{document}

\title{Exploring the vibrational series of pure trilobite Rydberg molecules}
\author{Max Althön}
\affiliation{Department of Physics and Research Center OPTIMAS, Rheinland-Pfälzische Technische Universität Kaiserslautern-Landau, 67663 Kaiserslautern, Germany}

\author{Markus Exner}
\affiliation{Department of Physics and Research Center OPTIMAS, Rheinland-Pfälzische Technische Universität Kaiserslautern-Landau, 67663 Kaiserslautern, Germany}

\author{Richard Blättner}
\affiliation{Department of Physics and Research Center OPTIMAS, Rheinland-Pfälzische Technische Universität Kaiserslautern-Landau, 67663 Kaiserslautern, Germany}

\corauthor{Herwig Ott\orcidlink{0000-0002-3155-2719}}{ott@physik.uni-kl.de}
\affiliation{Department of Physics and Research Center OPTIMAS, Rheinland-Pfälzische Technische Universität Kaiserslautern-Landau, 67663 Kaiserslautern, Germany}

\date{July 4, 2023}

\begin{abstract}
We report on the observation of two vibrational series of pure trilobite rubidium Rydberg molecules. They are created via three-photon photoassociation and lie energetically more than 15 GHz below the atomic 22$F$ state of rubidium. In agreement with theoretical calculations, we find an almost perfect harmonic oscillator behavior of six vibrational states. We show that these states can be used to measure electron-atom scattering lengths for low energies in order to benchmark current theoretical calculations. The molecules have extreme properties: their dipole moments are in the range of kilo-Debye and the electronic wave function is made up of high angular momentum states with only little admixture from the nearby 22$F$ state. This high-$l$ character of the trilobite molecules leads to an enlarged lifetime as compared to the 22$F$ atomic state. The observation of an equidistant series of vibrational states opens an avenue to observe coherent molecular wave-packet dynamics.
\end{abstract}

\maketitle

\section{Introduction}
Creating controllable molecules at ultralow temperatures offers a pathway to engineered ultracold quantum chemical reactions \cite{Carr_2009,doi:10.1126/science.aay9531, doi:10.1126/sciadv.aaq0083, doi:10.1126/science.1184121} and tests of fundamental physics and symmetries \cite{Carr_2009}. 
Molecules that possess sizeable electric dipole moments can be controlled by external electric fields making them candidates for quantum information processing \cite{PhysRevLett.88.067901,PhysRevLett.87.037901} and the production of strongly correlated many-body systems \cite{dipolar_quantum_gases,quantum_ferrofluid}. For dipolar molecules with multiple vibrational states electric field pulses have been proposed to create superposition states \cite{trilobite_wave_packet} and observe coherent wave-packet dynamics.   

Ultralong-range Rydberg molecules (ULRMs) \cite{Greene2000, ULRM_review, Eiles_2019} are a platform for creating such dipolar molecules in ultracold environments. In these molecules a neutral ground state atom is trapped inside the giant electronic wavefunction of a Rydberg state by a binding mechanism stemming from the electron-ground state scattering interaction. 
ULRMs have been found to be an ideal testbed for low-energy electron-ground state scattering \cite{PhysRevLett.123.073003,Boettcher_2016_mixed_singlet_triplet,internal_quantum_reflection, Cs_singlet_scattering} and could be used for the investigation of diabatic coupling schemes in molecules \cite{vibronic_interactions}. They can also be used as a starting point for the creation of ultracold anions \cite{Heavy_Rydberg}. 
Homonuclear ULRMs can have a permanent electric dipole moment due to the distinguishability of the ground state and Rydberg electron \cite{homonuclear_dipole_moment}.
For ULRMs corresponding to low-$l$ ($S$, $P$, $D$) Rydberg states this can reach about one Debye.
There are also two classes of molecules emerging from the mixing of multiple high-$l$ Rydberg states. These so-called butterfly \cite{butterfly} and trilobite molecules can have dipole moments on the order of kilo-Debye \cite{Cs_trilobite}, which are in special cases even larger than the bond length \cite{butterfly}.

Due to the high-$l$ nature of their electronic wave function trilobite molecules are in general not accessible with standard one- or two-photon photoassociation. Nevertheless, states with significant trilobite admixture have been produced via two-photon excitation both in Cs \cite{Cs_trilobite} and Rb \cite{Rb_trilobite}. In Cs the almost integer quantum defect of $S$ states leads to a mixing with the high-$l$ states, whereas in Rb a sizable admixture only exists for a specific principal quantum number, where the splitting between the $S$ state and the high-$l$ manifold matches the ground state hyperfine splitting. 
\newsavebox{\tempbox}
\begin{figure}[!hp]
\sbox{\tempbox}{\includegraphics[width=0.5\linewidth]{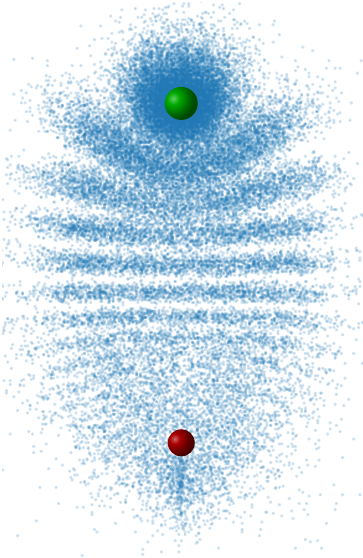}}
\hspace*{-0.32\linewidth}
\subfloat[]{\vbox to \ht\tempbox{%
  \vfil
  \includegraphics[scale=1.08]{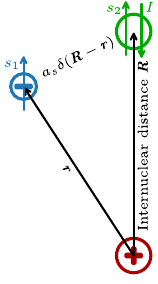}
  \vfil}\label{fig:1a}}
  \hspace*{-0.2\linewidth}
\subfloat[]{\usebox{\tempbox}\label{fig:1b}}\\
\caption{\footnotesize{a) Sketch of a Rydberg molecule. The coordinates of the Rydberg electron (blue) and ground state atom (green) relative to the Rydberg core (red) are denoted with black arrows. The relevant spins are that of the Rydberg electron $s_1$, the electron of the ground state atom $s_2$ and the nuclear spin of the ground state atom $I$. b) Sketch of a trilobite molecule. The Rydberg core and the ground state atom are shown (with exaggerated size) as red and green spheres respectively. The electronic probability density projected to 2D is indicated by the density of blue dots.}}
\label{fig:1}
\end{figure}

Here, we use three-photon excitation to produce pure Trilobite molecules in Rb over a wide range of frequencies and characterize their binding energies, lifetimes and dipole moments. We observe two vibrational series which are energetically split because of different angular momentum couplings and show that their lifetimes exceed that of the adjacent $22F$ state. Even for this relatively low principal quantum number, we find kilo-Debye dipole moments.

\section{Results}

ULRMs form due to the elastic scattering interaction of the Rydberg electron with a neutral ground state atom. To describe the scattering process, Fermi pseudo potentials \cite{Fermi, Omont} with energy-dependent scattering lengths are used. In atomic units, the interaction is given by

\begin{equation}
\begin{aligned}
\hat{V} = & A \hat{\vectorbold*{s_2}} \cdot \hat{\vectorbold*{I}} + \sum_{S,T} 2\pi \hat{\mathbb{P}}_{S,T} a_s^{S,T}(k) \delta\left( \vectorbold*{R} - \vectorbold*{r} \right) \\
&+ 6\pi \hat{\mathbb{P}}_{S,T} \left(a_p^{S,T}(k)\right)^3   \delta\left( \vectorbold*{R} - \vectorbold*{r} \right) \cev{\nabla} \cdot \vec{\nabla},
\end{aligned}
\label{eq:1}
\end{equation}
where $\vectorbold*{r}$ is the position of the Rydberg electron and $\vectorbold*{R}$ is the internuclear axis between the Rydberg core and the ground state atom, as shown in Fig.\,\ref{fig:1a}.  
The s- and p-wave scattering lengths  $a_{s/p}^{S,T}$ depend on the spins of the electrons resulting in singlet and triplet channels with the according projection operators $\mathbb{P}_{S,T}$. To explain the observed spectra the hyperfine interaction of the ground state atom $A \hat{\vectorbold*{s_2}} \cdot \hat{\vectorbold*{I}} $ needs to be taken into account. The scattering interaction depends on the Rydberg electron's momentum $k$ relative to the ground state atom, which is calculated semi-classically for every internuclear distance $R$ as $k = \sqrt{-1/n^2 + 2/R}$ (in atomic units). For the $k$-dependence of the singlet scattering lengths we use data provided by I. Fabrikant \cite{Fabrikant_1986, Bahrim_2001}. 
For the triplet channels we employ a model potential consisting of a polarization potential with an inner hard wall at variable distance from the ground state atom which captures the short-range physics \cite{PhysRevLett.123.073003, https://doi.org/10.1002/cphc.201600932}. By varying the position of the hard wall the scattering interaction can be tuned. 

\begin{figure}
\centering
\includegraphics[scale=1.08]{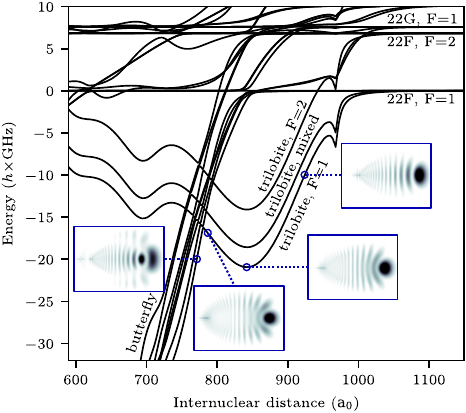}
\caption{\footnotesize{Born-Oppenheimer potential energy curves resulting from the electron-ground state scattering. The energy zero is the asymptotic pair state energy of a 22$F_{7/2}$ state and an $F=1$ ground state atom. The three labeled trilobite potentials result from different couplings of the spins $s_1$, $s_2$ and $I$. While the curves labeled $F=1$ and $F=2$ exhibit triplet scattering, the mixed curve has a mixture of singlet and triplet scattering, which in turn means a superposition of $F=1$ and $F=2$ in the ground state atom. The kink at about \SI{970}{\bohr} is due to the semi-classical calculation of the electron momentum $k$. Insets: Electronic probability density for the trilobite molecule shown at three different internuclear distances. At about \SI{780}{\bohr} the trilobite curves are crossed by the butterfly potentials, which result from a shape resonance in the p-wave scattering. The leftmost inset shows the electronic structure of a butterfly molecule.}}
\label{fig:potential}
\end{figure}

We diagonalize the Hamiltonian given in \cite{PhysRevA.95.042515}, which includes spin-orbit coupling of the p-wave scattering, at each internuclear distance. We consider a finite basis set consisting of two hydrogenic manifolds below the state of interest ($n=22$) and one manifold above it. The resulting Born-Oppenheimer potential energy curves are shown in Fig.\,\ref{fig:potential}. The energy curves belong to different types of molecule and show avoided crossings where the molecular character changes. Of particular interest for this work is the crossing between the trilobite and butterfly curves resulting in three mutually shifted potential wells which support multiple vibrational states.
While the lower potential well can be assigned to the $F=1$ ground state and triplet s-wave scattering, the middle potential curve consists of a mixture of the two hyperfine states and shows both singlet and triplet s-wave scattering \cite{Anderson_2014_angular_momentum_couplings, Boettcher_2016_mixed_singlet_triplet, Niederpruem_2016_spin_flips}. This mixture is due to the interplay between the hyperfine interaction and the electron scattering interaction, as both depend on the spin state of the ground state electron. The upper potential well for the $F=2$ ground state cannot be excited in our experiment, as we prepare the sample in the $F=1$ state. The bound states are then calculated from the potential curves with a shooting method by analyzing the density of states when varying the inner boundary condition \cite{internal_quantum_reflection}. 

To photoassociate the trilobite Rydberg molecules in the lower two wells we use a three-photon setup with lasers at \SI{780}{nm}, \SI{776}{nm} and \SI{1288}{nm}. This allows us to couple to the $22F$ state, which makes up about \SI{3}{\percent} of the electronic state. The first two lasers are blue detuned to the intermediate states  (5$P_{3/2}$ and 5$D_{5/2}$). The three-photon Rabi frequency is $2\pi\times\SI{250}{kHz}$. Our sample consists of $^{87}$Rb atoms in the $F=1$ ground state prepared in an optical dipole trap at \SI{1064}{nm}. The peak density is \SI{4e13}{ cm^{-3}} and the temperature is about \SI{150}{\micro K}. Per experimental cycle we perform 800 excitation pulses of \SI{1}{\micro s} duration. Before every excitation pulse the dipole trap is switched off. To detect the Rydberg excitations an extraction field is switched on after the excitation pulse and after a variable delay time a CO$_2$ laser pulse ionizes all Rydberg states. The resulting ions are guided via a reaction microscope \cite{remi1,remi2,remi3} to a space- and time-resolved multi-channel plate detector. This allows us to measure the momentum of the Rydberg core prior to ionization. Note that the recoil upon ionization with the CO$_2$ laser is negligible. 

Because of the large dipole moments of the trilobite molecules, precise electric field compensation during the excitation pulses is necessary. 
To achieve this we use the momentum imaging capabilities of our reaction microscope. In this field compensation measurement, the atoms are ionized and accelerated in the residual electric field for a variable wait time. Afterwards, we measure the momenta of the ions and extract the electric field from the linear dependence on the wait time. 

\begin{figure*}
\centering
\includegraphics[scale=1.2]{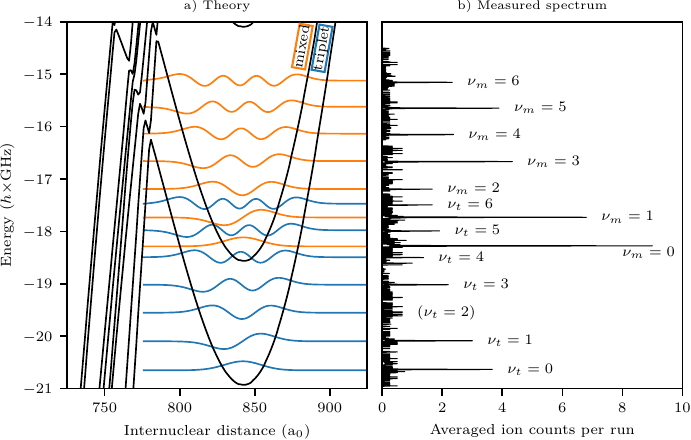}
\caption{a) Theoretical potential energy curves for a triplet s-wave asymptote of $a_s^T(k=0)=\SI{-14.2}{\bohr}$ and the resulting vibrational wavefunctions. The wavefunctions are drawn at their respective binding energies. b) Measured spectrum. The energy axis is shared with the theoretical potential shown on the left. The peaks are labeled with the vibrational quantum numbers $\nu_m$ for the mixed potential and $\nu_t$ for the triplet F=1 potential.}
\label{fig:spectrum}
\end{figure*}

\begin{figure}
\centering
\subfloat[]{\includegraphics[scale=1.08]{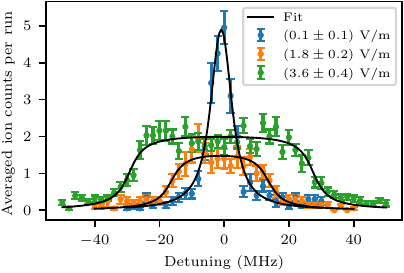}
\label{fig:dipolemoment_a}}\\
\subfloat[]{\includegraphics[scale=1.08]{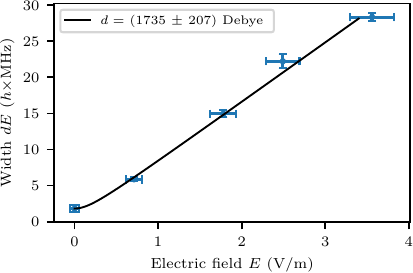}
\label{fig:dipolemoment_b}}
\caption{Measurement of the dipole moment for one of the trilobite molecules. a) Spectroscopy of the $\nu_m=0$ molecule at \SI{-18.274}{GHz} for different electric fields. The differences in peak heights are due to fluctuations in the sample preparation. The data is fitted by the convolution of a Lorentzian with a step function of width $2dE$ (see text). b) The resulting values of $dE$ from the fit of the spectrum are plotted against the electric field. The data is fitted to extract the dipole moment by the function $d\sqrt{E^2+E_0^2}$ allowing for an offset field $E_0$, which is on the order of \SI{0.1}{V/m}.}
\label{fig:dipolemoment}
\end{figure}

Fig.\,\ref{fig:spectrum} shows the molecular spectrum red detuned to the 22$F_{7/2}$ state covering the two trilobite potential wells. We observe a vibrational series of six bound states in each of the potential wells, which are equally spaced. The anharmonicity is less than 10 percent, confirming the harmonic oscillator shape of the potential wells. Thereby, the position of the highest vibrational state coincides with the crossing of the trilobite and the butterfly state, confirming that the well depth for both series is appropriately captured by theory.

Next, we analyze in detail the position of the vibrational states and the conclusions one can draw for the molecular potential. Inspecting the different terms in Eq.\,\ref{eq:1} shows that the molecular potential for the trilobite curve is directly proportional to the respective scattering length. High precision molecular Rydberg spectroscopy is therefore a tool to determine the electron-atom scattering lengths. Because of the crossing with the butterfly curves, both the triplet p-wave as well as the dominant triplet s-wave scattering channels have to be considered. Since the mixed trilobite has a small singlet admixture, the splitting of the two potential curves also depends on the singlet s-wave scattering length. Singlet scattering lengths calculated by a two-active-electron model \cite{Fabrikant_1986, Bahrim_2001} fit the observed splitting well. 
For the triplet p-wave scattering we find the $J=0$ shape resonance energy at $\SI{24.7\pm0.5}{meV}$, which agrees with the measurement of Engel et al.\ \cite{PhysRevLett.123.073003} within the error limits. The large margin of error is due to the relative insensitivity of the trilobite states to changes in the p-wave scattering.
For the more prominent triplet s-wave scattering we find a value of $a_s^T(k=0.0175)=\SI{-7.75\pm0.03}{\bohr}$ at the position of the potential minimum. Using the model potential to extrapolate this result to zero momentum yields $a_s^T(k=0)=\SI{-14.2}{\bohr}$. This asymptote differs significantly from previous experimental values (\SIrange{-15.2}{-16.1}{\bohr}) \cite{PhysRevLett.123.073003,Boettcher_2016_mixed_singlet_triplet,internal_quantum_reflection} measured at $k$ values near zero.
Given the high precision of the presented measurement, which is due to binding energies on the order of \SI{10}{GHz}, this points to an incorrect $k$ dependency of the scattering length as calculated from the model potential. We note that previous ab initio calculations \cite{Fabrikant_1986, Bahrim_2001} cannot explain the measured spectrum and therefore do not present an alternative. To resolve this, measurements of trilobite spectra at different principal quantum numbers can be used to probe different ranges of the electron momentum and thus present an opportunity to map out this dependency. With such measurements one can also test whether the semiclassical calculation of $k$ plays a role in the discrepancy. In fact, if the actual electron momentum is assumed to be about \SI{10}{\percent} larger than the semiclassical calculation, the binding energy of the triplet trilobite can be brought into line with the previously measured scattering length asymptote of Engel et al.\ \cite{PhysRevLett.123.073003}. Therefore, these exotic molecules could lead to a better theoretical understanding of the more general process of electron-atom scattering. 

A peculiar property of trilobite Rydberg molecules is their large permanent electric dipole moment. This stems from the large concentration of the electron density at the position of the ground state atom (see Fig.\,\ref{fig:1b}). The dipole moments are measured by applying an electric field and observing the broadening of the molecular line. As the rotational splitting cannot be resolved in our experiment, we fit the spectra with the convolution of a Lorentzian with a step function of width $2dE$ \cite{Cs_trilobite}. 
 From the fitted widths $dE$ for different electric fields we can deduce the dipole moment as shown in Fig.\,\ref{fig:dipolemoment}. We find electric dipole moments up to 1735 Debye, which corresponds to 0.8 times the internuclear distance. This reflects the highly efficient binding mechanism, which accumulates the electron density at the location of the ground state atom.
For the theoretical calculation of the dipole moments we write the electronic wavefunction at internuclear distance $R$ in the basis of the unperturbed states
\begin{equation}  
\ket{\Psi_\text{mol}^{(R)}} = \sum_i c_i^{(R)} \ket{i}
\end{equation}
and then integrate over the vibrational wavefunction $\Phi$
\begin{equation}
\langle d \rangle = \int  |\Phi(R)|^2 \sum_{i,j} c_i^{(R)*} c_j^{(R)}  \bra{i}\hat{d}\ket{j}    \text{d}R
\end{equation}

The experimental and theoretical results for selected vibrational states are presented in Table \ref{table}.
Experiment and theory are in good agreement, however, the experimentally determined dipole moments are systematically 10 - 15 \% larger than the theoretical values. This could be due to an unidentified systematic measurement error or incorrect theoretical dipole matrix elements as small systematic deviations add up due to the many states that contribute to the trilobite wave function. 

\begin{table}
\centering
\caption{Dipole moments and lifetimes of the trilobite rubidium Rydberg molecules. The binding energy is given with respect to the energy of the $22F_{7/2}$ state. The theoretical values are calculated with a triplet s-wave asymptote of $a_s^T(k=0)=\SI{-14.2}{\bohr}$. The theoretical radiative lifetime of the $22F_{7/2}$ state is \SI{6.3}{\micro \second}. A theoretical prediction of the lifetimes of the molecular states is beyond the scope of this paper.}
\begin{tblr}{cells={valign=m,halign=c}}
&  \SetCell[c=2]{c} Dipole moment (Debye) & & Lifetime (\si{\micro\second}) \\
 & exp. & theo. & exp. \\
$22F_{7/2}$ &  & &  $5.7\pm0.2$ \\
\hline
 $\nu_m=6$ &&  1472& $7.5\pm0.5$ \\
$\nu_m=5$ & $1632\pm193$ & 1477&$11.0\pm0.9$  \\
$\nu_m=4$ & & 1484&  \\
$\nu_m=3$ &  & 1489& $13.1\pm1.0$ \\
 $\nu_m=2$ &  & 1495&  \\
 $\nu_m=1$  && 1500&  \\
 $\nu_m=0$ & $1735\pm207$ &1507 & $14.6\pm1.0$ \\
\hline
 $\nu_t=6$ & &1472&  \\
 $\nu_t=5$  & &1477& $8.2\pm0.9$ \\
$\nu_t=4$  && 1482&  \\
 $\nu_t=3$  && 1489& $10.6\pm1.0$ \\
 $\nu_t=2$  && 1495&  \\
 $\nu_t=1$ & $1667\pm198$ & 1500&  \\
 $\nu_t=0$ & $1703\pm203$& 1505 & $12.0\pm0.9$ \\

\end{tblr}

\label{table}
\end{table}

The second important characteristic of Rydberg molecules is their lifetime. They reflect the different available decay channels, such as spontaneous emission, black-body induced transitions and molecular decay via tunneling towards shorter internuclear distances, leading to $l$-changing collisions or to associative ionization. The lifetimes are measured by varying the delay time between the excitation and ionization. We then count the number of ions that have zero momentum. This way, we also account for $l$-changing collisions, which result in an ion but come along with a large momentum \cite{State_changing}. Another possible outcome is associative ionization resulting in Rb$_2^+$ which can be distinguished by its larger TOF. Associative ionization as well as $l$-changing collisions stem from a tunneling process out of the potential well into the butterfly potential curve. 

The measured lifetimes are given in Table \ref{table}. We first note that even the shortest measured lifetime is larger than the lifetime of the atomic $22F$ Rydberg state and the ground state in each potential well has double its lifetime. This reflects the multitude of involved high-$l$ states in the trilobite molecules, resulting in slower radiative decay. As expected, the tunneling processes are more prominent for higher vibrational states, resulting in a shorter lifetime than the deeply bound vibrational ground states. This is corroborated by a roughly 60 \% increase in rates for both $l$-changing collisions and associative ionization when comparing $\nu_m=6$ with the ground state. However, the quantitative dependence on the vibrational quantum number needs further investigation, as e.g.\ vibronic coupling effects \cite{vibronic_interactions} lead to a superposition of butterfly and trilobite states and thus influence the decay dynamics.

\section{Conclusion and Outlook}
We have measured two vibrational series of pure trilobite Rydberg molecules by employing three-photon photoassociation. With this method the creation of trilobite molecules in any element that has a negative s-wave scattering length should be possible, as the quantum defects for the admixed atomic state (here $F$-state) are rather small and the coupling with the trilobite state is sizable. 
We find kilo-Debye dipole moments and lifetimes longer than the coupled atomic state. The observed spectra can be theoretically explained by adjusting the triplet s-wave scattering length. While the resulting agreement is excellent, the extrapolated scattering length asymptote disagrees with previous measurements and merits further theoretical and experimental work. As a logical next step one can extend the measurements for different principal quantum numbers and thus probe different ranges of electron momenta. This allows to map out the scattering length dependence on $k$. The discrepancy found in the scattering length asymptote might also be due to the semi-classical treatment of the electron momenta and an extended measurements series might support or discard this explanation. 

Additionally, for higher principal quantum numbers vibronic coupling effects between the trilobite and butterfly curves become more pronounced and these molecules could serve as a benchmark for theoretical calculations \cite{vibronic_interactions}. It has also been predicted that at certain principal quantum numbers conical intersections essentially stop the $l$-changing collision processes \cite{conical_intersections}, which could be checked with our reaction microscope. Finally, the shape of the potential well is suitable to study coherent wave packet dynamics. Note that the quality factor of the potential well is $3\times 10^4$. Using ns and ps laser pulses in a pump probe scheme provides the required time resolution for such experiments.

\section{Acknowledgements}
We would like to thank Frederic Hummel, Peter Schmelcher and Matt Eiles for helpful discussions. This project is funded by the Deutsche Forschungsgemeinschaft (DFG, German Research Foundation) – project number 316211972 and 460443971.

\section{Author contributions}
M.A., M.E., R.B. performed the experiments. M.A. and M.E. analyzed the data. M.A. performed the theoretical calculations and prepared the initial version of the manuscript. H.O. conceived and supervised the project. All authors contributed to the data interpretation and manuscript preparation. 
\section{Data availability}
The data that support the findings of this study are available from the corresponding
author upon reasonable request.
%

\end{document}